\documentclass[10pt,a4paper]{article} 
\usepackage{amsthm}
\usepackage{hyperref}
\title{Duality between Lines and Points}

\author{Sanjeev Saxena\thanks{E-mail: ssax@iitk.ac.in}\\
Dept. of Computer Science and
Engineering,\\ Indian Institute of Technology,\\
Kanpur, INDIA-208 016}

\date{\today}
\newtheorem{observation}{Observation}
\newtheorem{lemma}{Lemma}
\newtheorem{cor}{Corollary}
\begin{document}
\maketitle

\subsection*{\centering{Abstract}}

There are several notions of duality between lines and points. In this
note, it is shown that all these can be studied in a unified way. Most
interesting properties are independent of specific choices.

It is also shown that either dual mapping can be its own inverse or it
can preserve relative order (but not both).

Generalisation to higher dimensions is also discussed. An elementary
and very intuitive treatment of relationship between arrangements in
$d+1$ dimensions and searching for $k$-nearest neighbour in
$d$-dimensions is also given.

\section{Dual Mapping}

A non-vertical line is determined by two parameters--- e.g., $m$ and
$c$ in $y=mx+c$ or $a$ and $b$ in $\frac{x}{a}+\frac{y}{b}=1$. And a
point in 2-d also requires two parameters (coordinates, e.g., $x$ and
$y$ in Cartesian and $r$ and $\theta$ in polar).

Various authors suggest different ways to map points to lines and
vice-versa
\begin{description}
\item[Ja'Ja'\cite{J},Lee and Ching\cite{LC}] point $(r,s)$ is mapped to line $y=rx+s$ and
line $y=mx+c$ is mapped to point $(-m,c)$

\item[O'Rourke\cite{R}] Suggests two mappings first maps line
$y=mx+c$ to point $(m,c)$ and conversely.

The other which he actually uses is maps point $(r,s)$ to line
$y=2rx-c$ and conversely.

\item[Berg et.al.\cite{B}] point $(r,s)$ is mapped to line $y=rx-s$
and dual of line $y=mx+c$ is the point $(m,-c)$.

\item[Chazelle, Guibas and Lee\cite{CGL}] point $(r,s)$ is mapped to line
$rx+sy+1=0$ and conversely.
\end{description}

Let us look at general dual mapping in which point $(r,s)$ is mapped
to line $y=\alpha rx+\beta s$. And a line $y=mx+c$ gets mapped to
point $(\mu m,\lambda c)$.

\begin{observation}\cite{R}
There is a one-to-one correspondence between all non-vertical lines
and all points in the plane.
\end{observation}

Remark: This property will not be true for more general ``dual'' lines
like $y=(ap+bq)x+(cp+dq)$.

We will like the following property to be true for the mapping:
\begin{quote}
If a point $p$ lies on a line $L$, then dual of line $L$, say $d(L)$
(is a point) $d(L)$ which should lie on dual of point $p$, say $d(p)$
(which is a line $d(p)$).
\end{quote}

A general line through point $(r,s)$ will be $y-s=m(x-r)$. This line
gets mapped to a point $(\mu m, \lambda(s-mr))$. As this point must
lie on line $y=\alpha rx+\beta s$, we get
\begin{eqnarray*}
\lambda(s-mr) &=& \alpha (\mu m)+\beta s \mbox{ or}\\
s(\lambda-\beta) &=& m(\alpha\mu +r \lambda \mbox{ or}\\
\lambda &=& \beta \mbox{ and }\\
\alpha\mu &=&- \lambda \mbox{ or}\\
\alpha\mu &=& -\beta
\end{eqnarray*}

Thus, the {\em{permissible transforms}} will map \\
point $(r,s)$ to line
$y=\alpha (rx -\mu s)$ and will map\\
line $y=mx+c$ to point $(\mu m,
-\alpha \mu c)$.

\begin{lemma}
If $d(~)$ is a permissible dual transform, then a point $p$ lies on a
line $L$, then dual of line $L$, $d(L)$ lies on $d(p)$, the dual of
point $p$.
\end{lemma}

\begin{cor}\cite{R}
Two lines $L_1$ and $L_2$ intersect in a point $P$, iff, dual $d(P)$
passes though $d(L_1)$ and $d(L_2)$.
\end{cor}

As we have two free parameters we can impose additional ``desirable''
conditions. Two of popular desirable conditions are
\begin{enumerate}
\item If dual of point $p$ is line $L$, then dual of line $L$ should
be point $p$.
\item If point $p$ lies above line $L$, then dual of $p$ should lie
above dual of $L$.
\end{enumerate}

Let us first look at first condition. Dual of line $y=(\alpha
r)x+(-s\mu \alpha)$ will be point $(\mu (\alpha r),(-\mu
\alpha)(-s\mu\alpha))$. For this point to be $(r,s)$ we need
$\mu\alpha=1$.

\begin{lemma}
If $\alpha\mu=1$, then dual $d$ is its own inverse.
\end{lemma}

Remark: These transforms have only one free parameter. Point $(r,s)$
gets mapped to line $y=\alpha rx -s)$ and line $y=mx+c$ gets mapped to
the point $(\frac{1}{\alpha} m, -c)$.

Let us now look at the second condition. A point $(r,s)$ is above line
$y=mx+c$ if $s-rm>c$ or $0>c+rm-s$ or $c+rm-s<0$.

Dual of point $(r,s)$ is the line $y=\alpha (rx-\mu s)$ and dual of
line $y=mx+c$ is the point $(\mu m, -\alpha \mu c)$. Line 
$y=\alpha (rx-\mu s)$ is above the point $(\mu m, -\alpha \mu c)$, if
$$(-\alpha \mu c)-\alpha r(\mu m)< -\alpha \mu s$$
or $$-(\alpha \mu)(c+rm-s)<0$$ As $c+rm-s<0$, $-(c+rm-s)>0$ or
$\alpha\mu<0$.

\begin{lemma}
If $\alpha\mu<0$, then if a point $p$ lies above line $L$ then $d(p)$
lies above $d(L)$. And if

If $\alpha\mu>0$, then if a point $p$ lies above line $L$ then $d(p)$
lies below $d(L)$ (i.e. order gets reversed).
\end{lemma}

\begin{observation}
Both Conditions can not be simultaneously satisfied.
\end{observation}

If we define {\em{vertical distance}}\cite{D} between a point $(r,s)$ and line
$y=mx+c$ to be $|mr+c-s|$, then, from the proof of lemma, in the dual
space the vertical distance gets scaled by $|\alpha\mu|$. 

Parallel lines $y=mx+c$ and $y=mx+b$ will be mapped to points $(\mu m,
-\alpha \mu c)$ and $(\mu m, -\alpha \mu b)$. Or (vertical) distance
between these two points is $|\alpha\mu(b-c)|$.

Thus, if wish vertical distance to be preserved, then $\alpha\mu=\pm
1$.

Har-Peled \cite[Exercise 31.2]{D} observes that no duality can
preserve exactly orthogonal distances between points and lines.

\section{Applications}

Assume $S$ is a set of points in the plane.  A point $p$ is on convex
hull, iff there is a non-vertical line $L$ through $p$ s.t., all
points of $S$ are below the line $L$. 

For each point of $S$ (including $p$), there is a line in dual space.
Dual of $L$ is a point which lies on line $d(P)$. All dual lines
$d(S)$ should be on one side of $d(L)$. Or $d(L)$ lies on (upper or
lower, depending on sign of $\alpha \mu$) envelope of lines $d(S)$.

Let us consider intersection of half-planes. Each half plane is either
$y\leq m_i x+c_i$, which will be called {\em{top constraint}} or
$y\geq m_ix+c_i$, which will be called {\em{bottom constraint}}.

The ``feasible region'' for top (respectively, bottom) constraints
will be a convex region. Assume point $(x_0,y_0)$ is on the boundary
of this convex region. As it satisfies all top constraints it is below
lines $y=m_ix+c_i$ (if $i$ is top constraint). And as it is on the
boundary, it is on at least one such line. In dual space, point
$(x_0,y_0)$ gets mapped to a line (say $L_0$), and each line
$y=m_ix+c_i$ will get mapped to a point (say $p_i$). Moreover, all
points $p_i$ in dual space, will be on one side of the line $L_0$ and
at least one point will be on this line. Thus, line $L_0$ will be
tangent to the convex hull.

An edge $e$ in the convex hull (in dual space) is a line between two
dual points, say $d(L_1)$ and $d(L_2)$. This edge $e$ will correspond
to a point (say $X$) in the untransformed domain. As (dual) line
$d(X)$ (which contain segment $e$) passes through points $d(L_1)$ and
$d(L_2)$, point $X$ will lie on lines $L_p$ and $L_q$, i.e., point $X$
will be the point of intersection of these two lines.

Thus, line segments of convex hull define define extreme point of
feasible solution space, and both sets occur in the same relative
order. Let us assume we have determined the convex hull.

We can similarly, deal with bottom constraints $B$. As both these
hulls are sorted (say on $x$ coordinate), these can be merged in
linear time. The merged sets define a vertical slab, in which there
are at most two boundary segments, one from lower and one from upper.
Hence, we can determine the boundary in linear time.

In linear programming, assume we wish to maximise $cx+dy$. As
each corner point $(p_i,q_i)$ of solution space, we find
$V_i=cp_i+dq_i$ and find the largest $V_i$. This again takes linear
time, if the solution space is known.

Kernel of a polygon is the region from which entire polygon is
visible. We interpret each edge of polygon as a half plane (the side
containing the interior). Intersection of these regions will give the
kernel.

\section{Duality in $d$-dimensions}

Let us generalise dual maps used (see e.g.,\cite{D,E}). Assume point
$P=(p_1,{\ldots} ,p_d)$ is mapped to hyperplane $x_d=\sum_{i=1}^{d-1}
a_ip_ix_i+a_dp_d$ for some constants $a_i$. 
And hyperplane $L$ with equation $x_d=\sum_{i=1}^{d-1} m_ix_i+c$ is mapped
to point $(b_1m_1,b_2m_2,{\ldots} ,b_{d-1}m_{d-1},b_dc)$.

Edelsbrunner\cite[p4]{E} in first map takes all $a_i=1$ and in the second
\cite[p13]{E}
$a_i=2$ for $1\leq i\leq d-1$ and takes $a_d=-1$. In another map\cite[p17]{E},
point $P$ is mapped to hyper-plane $\sum_{i=1}^d p_ix_i=1$. 

If point $P$ lies on hyperplane $L$ then $p_d=\sum m_ip_i+c$.
For transformed point  $(b_1m_1,b_2m_2,{\ldots} ,b_{d-1}m_{d-1},b_dc)$
to lie on transformed plane, we must have
$$(b_dc)=\sum a_i(b_im_i)p_i+a_dp_d=\sum a_i(b_im_i)p_i+a_d\left(\sum
m_ip_i+c\right)=\sum m_ip_i(a_ib_i+a_d)+a_dc$$
Thus, $b_dc=a_dc$ or $b_d=a_d$ and $a_ib_i=-a_d$ or $b_i=-a_i/a_d$.

Or the general transforms are
\begin{description}
\item[Point] $P=(p_1,{\ldots} ,p_d)$ is mapped to hyperplane
$x_d=\sum_{i=1}^{d-1} a_ip_ix_i+a_dp_d$ for some constants $a_i$. 

\item[Plane] And hyperplane $L$ with equation $x_d=\sum_{i=1}^{d-1}
m_ix_i+c$ is mapped to point
$$\left(-a_d\frac{m_1}{a_1},-a_d\frac{m_2}{a_2},{\ldots}
,-a_{d}\frac{m_{d-1}}{a_{d-1}},a_dc\right)$$
\end{description}
Dual of point $\left(-a_d\frac{m_1}{a_1},-a_d\frac{m_2}{a_2},{\ldots}
,-a_{d}\frac{m_{d-1}}{a_{d-1}},a_dc\right)$ is the plane
$$x_d=\sum a_i\left(-a_d\frac{m_i}{a_i}\right)x_i+a_d(a_dc)=
-\sum a_dm_ix_i+a^2_dc$$
For this to be the original plane (dual to be its own inverse) $a_d=-1$
Thus, for this case, the transforms are:
\begin{description}
\item[Point] $P=(p_1,{\ldots} ,p_d)$ is mapped to hyperplane
$x_d=\sum_{i=1}^{d-1} a_ip_ix_i-p_d$ for some constants $a_i$. 

\item[Plane] And hyperplane $L$ with equation $x_d=\sum_{i=1}^{d-1}
m_ix_i+c$ is mapped to point
$$\left(\frac{m_1}{a_1},\frac{m_2}{a_2},{\ldots}
,\frac{m_{d-1}}{a_{d-1}},-c\right)$$
\end{description}

Let us now look at the second condition. A point $P$ is above the
plane $L$ if $p_d\geq \sum m_i p_i+c$ or
$$p_d- \sum m_i p_i- c\geq 0$$

Dual of $P$ is the hyper plane $x_d=\sum_{i=1}^{d-1}
a_ip_ix_i+a_dp_d$.

And dual of hyper plane $L$ is the point
$\left(-a_d\frac{m_1}{a_1},-a_d\frac{m_2}{a_2},{\ldots}
,-a_{d}\frac{m_{d-1}}{a_{d-1}},a_dc\right)$

Dual of point (hyperplane) is above the dual of plane (point) if
the dual point (of original hyper plane) is below dual plane (of
original point) thus:
$$ca_d \leq \sum a_ip_i \left(-a_d\frac{m_i}{a_i}\right)+a_dp_d
= a_d\left( p_d-\sum p_i m_i\right)$$
Or $$0\leq a_d \left( p_d-\sum p_i m_i-c\right)$$
As the term in the brackets is positive, it implies $a_d$ is also
positive, i.e., $a_d>0$. Thus, again, both conditions can not be true.

Parallel hyperplanes
$x_d=\sum m_ix_i+c$ and $x_d=\sum m_ix_i+b$ will get mapped to points
$$\left(-a_d\frac{m_1}{a_1},-a_d\frac{m_2}{a_2},{\ldots}
,-a_{d}\frac{m_{d-1}}{a_{d-1}},a_dc\right)$$ and
$$\left(-a_d\frac{m_1}{a_1},-a_d\frac{m_2}{a_2},{\ldots}
,-a_{d}\frac{m_{d-1}}{a_{d-1}},a_db\right)$$
Thus, the vertical distance $|c-b|$ gets scaled to $|a_d(c-b)|$ to
preserve vertical distance $a_d=\pm 1$ (as before).

Remark: As first $d-1$ coordinates of dual points are identical, other
distances are not preserved.

Vertical distance between point $(p_1,{\ldots} ,p_d)$ and hyperplane
$x_d=\sum_{i=1}^{d-1} m_ix_i+c$ is $\left|p_d-\sum_{i=1}^{d-1} m_ix_i-c \right|$
and between dual hyperplane $x_d=\sum_{i=1}^{d-1} a_ip_ix_i+a_dp_d$ and
dual point $-a_d(m_1/a_1,{\ldots} ,m_{d-1}/a_{d-1},-c)$ is
$$\left|a_dc+a_d \sum_{i=1}^{d-1} a_ip_i(m_i/a_i)-a_dp_d\right|=
|a_d|\left|c+\sum_{i=1}^{d-1} p_i m_i- p_d\right|$$
Thus, again distances will be preserved if $|a_d|=1$ or $a_d=\pm 1$.

\subsection{Application to $k$-Nearest Neighbours and Arrangements}

An elementary and very intuitive treatment of relationship between
arrangements in $d+1$ dimensions and searching for $k$-nearest
neighbour in $d$-dimensions is also given.

Assume $S=\left\{P_1,{\ldots} ,P_n\right\}$ is a set of $n$ points in
the $R^d$, the $d$-dimensional Euclidean space. Assume
$P=(p_1,{\ldots} ,p_d)$ and $Q=(q_1,{\ldots} ,q_d)$ are two points of
$S$.

A query (or test) point $X=(x_1,{\ldots} ,x_d)$ will be closer to $P$ then to
$Q$ iff
\begin{eqnarray*}
\sum_{i=1}^d (x_i-p_i)^2 &<& \sum_{i=1}^d (x_i-q_i)^2 \mbox{ or }\\
\sum_{i=1}^d \left({x_i^2}-2x_ip_i+p_i^2\right) &<& 
\sum_{i=1}^d \left({x_i^2}-2x_iq_i+q_i^2\right) \mbox{ or}\\
2 \sum_{i=1}^d x_iq_i - \sum_{i=1}^d q_i^2 
&<& 2 \sum_{i=1}^d x_ip_i - \sum_{i=1}^d p_i^2 
\end{eqnarray*}
For $R=(r_1,{\ldots} ,r_d)$ define $f(R)=2 \sum_{i=1}^d x_ir_i -
\sum_{i=1}^d r_i^2$.

Then above condition becomes, $f(P)>f(Q)$.

As equation $z=f(R)=2 \sum_{i=1}^d x_ir_i -
\sum_{i=1}^d r_i^2$
is an equation of $d+1$ dimensional hyper-plane, the above condition is
equivalent to saying that\\
hyper-plane $z=2 \sum_{i=1}^d x_ip_i - \sum_{i=1}^d p_i^2 $
is above the hyper-plane 
$z=2 \sum_{i=1}^d x_ip_i - \sum_{i=1}^d p_i^2 $

Remark: Instead of $z=f(R)$ we can also choose any monotonic function
(like $z^2=f(R),\sqrt{z}=f(R)$ etc.). But $z=f(R)$ appears to be the
simplest.

Hence, if we draw an arrangement of $n$-hyper-planes with plane
$z=f(P_i)$ the $i$th point, then the nearest
neighbour of query point $X$ will be the topmost hyper-plane above
$X$ (the one visible from $(x_1,{\ldots}x_d ,\infty)$). The second nearest
neighbour will be the second top most hyper-plane and so on. Hence, to find
$k$th closest point, we need to only consider hyper-planes which have
$(k-1)$ hyper-planes above them.

Edelsbrunner\cite[p23]{E} considers a map which transforms point
$P$ to $x_{d+1}=\sum_{i=1}^d 2p_ix_i-\sum_{i=1}^d p_i^2$.
Here $a_i=2$ for $i=1,{\ldots} ,d$ and $c=- \sum_{i=1}^d p_i^2 $

Thus, this can also be viewed as mapping a point in $d$-dimensions to
a ``special plane'' (e.g., one constrained to be tangent to a
paraboloid or a hyper-sphere) in $d+2$ dimensions.

\subsection*{\center{Acknowledgements}}

I wish to thank students who attended lectures of CS663 (2024-2025)
for their comments and reactions.

\bibliography{general}

\end{document}